\documentclass{article} 
\usepackage{latexsym}
\usepackage{graphics}
\usepackage{epsfig}
\usepackage[mathscr]{eucal}
\usepackage{amscd}
\usepackage{amssymb}
\usepackage{amsmath}
\usepackage{amssymb}
\usepackage[dvips]{color}

\newcommand{\re}{\mathbb R}

\newcommand{\cyl}{\mathrm{cyl}}

\newtheorem{lemma}{Lemma}
\newtheorem{claim}{Claim}
\newcommand{\qed}{\Box}
\newcommand{\fS}{\mathfrak S}
\newcommand{\fA}{\mathfrak A} 
 
\newcommand{\fSA}{{\mathfrak S}_\fA}

\newcommand{\cH}{\mathcal H}

\newcommand{\bC}{\mathbb C}
\newcommand{\cB}{\mathcal B}

\newcommand{\tmuv}{|{\muv}|}

\newcommand{\hC}{\widehat C}

\newcommand{\muv}{\mu_{\mathsf v}}
\newcommand{\nuv}{\eta_{\mathsf v}}
\newcommand{\bmuv}{\overline \muv}
\newcommand{\bnuv}{\overline \nuv}
\newcommand{\wmuv}{\tilde \muv}
\newcommand{\twmuv}{|\wmuv|}
\newcommand{\bD}{\overline D}
\newcommand{\ginf}{\gamma^{(\infty)}}

\newcommand{\tpsi}{\widetilde \psi}
\newcommand{\cn}{C^n}
\newcommand{\fu}{\mathrm{u}}
\newcommand{\un}{\mathrm{v}}
\newcommand{\GF}{\Gamma_2}

\newcommand{\tf}{\tilde f}
\newcommand{\tg}{{\tilde \gamma}}
\newcommand{\tG}{{\tilde \Gamma}}
\newcommand{\cw}{c_{\mathrm w}}
\newcommand{\ww}{\mathrm w}
\newcommand{\iS}{\mathcal I}
\title{On extending the Quantum Measure} 
\author{Fay Dowker${}^1$, Steven Johnston${}^1$, Sumati Surya${}^2$
\\ \\
\small ${}^1$ Blackett Laboratory, Imperial College, London, SW7
  2AZ, UK \\ 
\small ${}^2$ Raman Research Institute, Sadashivanagar, Bangalore, 560080, India \\ \\ 
} 
\begin{document}
{
\maketitle 
\begin{abstract}

 We point out that a quantum system with a strongly positive quantum measure or decoherence
  functional gives rise to a vector valued measure whose domain is the algebra of events or physical
  questions. This gives an immediate handle on the question of the extension of the decoherence
  functional to the sigma algebra generated by this algebra of events. It is on the
  latter that the physical transition amplitudes directly give the decoherence functional. Since the
  full sigma algebra contains physically interesting questions, like the return question, extending
  the decoherence functional to these more general questions is important. We show that the
  decoherence functional, and hence the quantum measure, extends if and only if the associated
  vector measure does.  We give two examples of quantum systems whose decoherence functionals do not
  extend: one is a unitary system with finitely many states, and the other is a quantum
    sequential growth model for causal sets.  These examples fail to extend in the
  formal mathematical sense and we speculate on whether the conditions for extension are
  unphysically strong.

\end{abstract}

\section{Introduction}

The need for a measurement independent interpretation of quantum theory is perhaps most keenly felt
when constructing a theory of quantum cosmology. In describing the physics of the very early
universe we are confronted with the dilemma of how to interpret the quantum formalism in the absence
of external measurements. The standard interpretation places an emphasis on the state vector, and
gives primary status to external measurements but it is questions of spacetime form that are of
interest in quantum cosmology {\textit e.g.} how likely is it for a homogeneous and isotropic
universe to arise from an initial big-bang type singularity?

Quantum Measure Theory \cite{qmeasure,qmeasuretwo,rob,qrw,qcovers,gudder} is a formulation of
quantum theory based on the path integral.  Since measuring devices play no fundamental role in this
approach, it is ideally suited to examining theories of quantum cosmology.  Quantum Measure Theory
gives conceptual primacy to the {\sl sample space} of histories or spacetime configurations,
$\Omega$, which is summed over in the path integral and takes its cue from the measure theoretic
formulation of classical stochastic dynamics.  A physical quantum system is described by a {\sl
  quantum measure space}, $(\Omega,\fA, \mu)$, where $\fA$ is an {\sl event algebra} or set of
propositions about the system and dynamical information is contained in the {\sl quantum measure}
$\mu: \fA \rightarrow \re^+$ which is given by the path integral.  $\mu$ obeys the {\sl quantum sum
  rule} \cite{qmeasure}
\begin{equation}\label{qsr}
\mu(\alpha \cup \beta \cup \gamma )= \mu(\alpha\cup \beta) +
\mu( \alpha\cup \gamma) + \mu(\beta \cup
\gamma) - \mu(\alpha) - \mu(\beta) -\mu(\gamma)
\end{equation}
all for pairwise disjoint sets $\alpha, \gamma,\beta \in \fA$.  $\mu$ is not in general a
probability measure (even with the normalisation $\mu(\Omega)=1$) since it does not satisfy the
Kolmogorov sum rule in the presence of quantum interference: $\exists \,\, \alpha,\beta \in \fA$
with $\alpha \cap \beta = \emptyset$ and $\mu(\alpha \cup \beta) \neq \mu(\alpha) + \mu(\beta)$.
Thus, Quantum Measure Theory is a genuine generalisation of classical stochastic dynamics.

In classical stochastic theories, the probability measure on the event algebra is often defined
indirectly in terms of transition probabilities from one momentary state to another. The classical
random walk is a standard example.  The transition probabilities define, directly, the probability
of events which are limited in time, for example: ``Is the walker at site $x$ at time $t$?" Given an
initial position at say $t=0$, one calculates the probabilities of each of the walks with $t$ steps
that end at $x$ using the transition probabilities and adds them together.  The probability of
certain physically interesting events, however, cannot be directly calculated in this way. These
questions involve arbitrarily long times and are epitomised by the return question: ``Does the
walker ever return to the origin?''  In order to find the probability of such events, one must be
able to extend the probability measure on the finite-time events to infinite-time events. For
non-negative measures this is guaranteed by the Carath\'eodory-Kolmogorov extension theorem, which
gives a unique extension of the measure on the algebra of finite-time events to the sigma algebra it
generates.  The sigma algebra is closed under countable unions and intersections and one can show
that the ``return event'' is an element of this algebra \cite{kac,halmos}.

Similarly, in most quantum systems the quantum measure derives from transition amplitudes from one
momentary state to another. To make predictions about infinite-time events, like the quantum
analogue of the return question, requires an extension of the quantum measure to an algebra that
includes such events. There is however no known analog of the Carath\'eodory-Kolmogorov extension
theorem for quantum measures. In this paper we take first steps in investigating this issue. The
technical development that helps this analysis is the histories Hilbert space construction from
quantum measures which derive from a decoherence functional \cite{hilbert}.  We show that the
quantum measure is equivalent to a derived {\sl {vector measure}}, {\sl {i.e.}} a measure valued in
this histories Hilbert space. Unlike the quantum measure, vector measures are additive and have been
studied extensively in the literature \cite{diesuhl}.  In this paper, we address the question of
extension of the quantum measure by studying the derived vector measure for a class of systems in
which the histories Hilbert space is finite dimensional.

In Section \ref{qvm} we define the quantum vector ``pre-measure'' on the algebra of finite-time
events after reviewing the histories Hilbert space construction of \cite{hilbert}. In Section
\ref{finite} we examine finite dimensional unitary systems which evolve in discrete time steps, and
show that for a generic evolution, the quantum vector pre-measure does not extend to a quantum
vector measure. In Section \ref{cp} we define the complex percolation sequential growth dynamics for
causal sets and show that the quantum vector pre-measure does not extend except when the amplitudes
are real and non-negative. In Section \ref{disc} we discuss the implications of our results. For the
complex percolation type models, the question of an extension is tied closely to the construction of
covariant observables. Using an example we show that the lack of an extension is related to other
pathologies.

\section{The Quantum Vector Measure} 
\label{qvm} 

In this section we show that a quantum measure space in which the quantum measure derives from a
strongly positive decoherence functional is equivalent to a vector measure which takes its values in
a Hilbert space.

In order to formulate quantum dynamics as a measure space, the sample space $\Omega$ is taken to be
the set of histories summed over in the path integral. For a single particle in $\re^3$ whose
evolution starts at some initial time this is the space of all trajectories that are infinite to the
future, while for a scalar field in $\re^3\times \re$ it is the set of all spacetime field
configurations.  An event algebra $\fA$ over the sample space $\Omega$ is a collection of subsets of
$\Omega$ that forms an {\sl algebra} or {\sl field of sets} over $\Omega$. Thus, (i) $\alpha \in \fA
\Rightarrow \alpha^c \in \fA $, where $\alpha^c$ is the complement of $\alpha$ in $\Omega$ and (ii)
$\alpha \cap \beta \in \fA$ and $\alpha \cup \beta \in \fA$ for any $\alpha, \beta \in \fA$.  A
sigma algebra $\fS$ satisfies, in addition to (i) and (ii), closure under countable unions.  The
sigma algebra $\fS_\fA$ generated by an algebra $\fA$ is defined to be the (unique) smallest sigma
algebra containing $\fA$.  In what follows, in order to distinguish between a measure on an algebra
and that on a sigma algebra, we will refer to the former as a {\sl pre-measure} and the latter as a
measure (thus, a pre-measure is a measure if the algebra on which it is defined is a sigma algebra).

A decoherence functional is a complex function $D: \fA \times \fA \rightarrow \bC$ which 
represents the quantum interference between two events. $D$ is
\cite{qrw}
\begin{enumerate}  
\item Hermitian: For all $\alpha, \beta \in \fA$, we have $D(\alpha,\beta )=  D^\ast (\beta,\alpha)$.  
\item 
Finitely bi-additive: For any $\alpha \in \fA$ and $m$ mutually disjoint $\beta_i \in \fA$, we have
$D(\alpha, \cup_{i=1}^m \beta_i)=\sum_{i=1}^m D(\alpha,\beta_i)$. Similarly, for any $\beta \in \fA$ and $m$ mutually disjoint $\alpha_i \in \fA$,
$D(\cup_{i=1}^m \alpha_i,\beta)=\sum_{i=1}^m D(\alpha_i,\beta)$.  
\item Normalised: $D(\Omega, \Omega) =1 $. 
 
\item Strongly positive: For any finite collection of $\{ \alpha_i \}$ in $\fA$,
the matrix $M_{ij}\equiv D(\alpha_i,\alpha_j)$ is positive semi-definite, i.e., it has non-negative eigenvalues.

\end{enumerate} 

The quantum pre-measure $\mu: \fA \rightarrow \re^+$ derives from the decoherence functional via
$\mu(\alpha) \equiv D(\alpha,\alpha)$ and we see that the biadditivity of $D$ means that $\mu$
satisfies the quantum sum rule (\ref{qsr}) but $\mu(\alpha\cup\beta) \neq \mu(\alpha) + \mu(\beta)$
if $Re(D(\alpha,\beta)) \neq 0$, for disjoint $\alpha$ and $\beta$.  In what follows we will refer
interchangeably to both $D$ and $\mu$ as the {\sl quantum pre-measure}\footnote{While $\mu$ has no
  standard measure theoretic analogue, since it is not additive, the decoherence functional belongs
  to the class of ``biadditive complex-valued pre-measures'', also called ``bi-measures'' or
  ``poly-measures'' \cite{bimeasure}.}.  The construction in \cite{hilbert} of a Hilbert space from
the event algebra $\fA$ and the decoherence functional $D$ implies that the quantum measure is
equivalent to a Hilbert space valued measure which {\sl{is}} additive, unlike the quantum
measure\footnote{We thank Rafael Sorkin for this observation.}. This gives us a useful ``vector
measure'' avatar of the quantum measure.  We now briefly review vector measures and pre-measures.

A {\sl vector pre-measure} \cite{diesuhl} $\nuv$ is a function from an
algebra $\fA$ over $\Omega$ to a Banach space $\cB$ which is {\sl
finitely additive}, i.e., for every disjoint pair $\alpha,\beta \in
\fA$
\begin{equation}
\nuv(\alpha \cup \beta) = \nuv(\alpha) + \nuv(\beta).   
\end{equation}     
If $\fS$ is a sigma algebra, a {\sl vector measure} $\bnuv:\fS
\rightarrow \cB$ is moreover required to be {\sl countably additive}
\begin{equation} \label{ucc} 
\bnuv\left(\bigcup_{n=1}^\infty \alpha_n\right)= \sum_{n=1}^\infty \bnuv(\alpha_n) 
\end{equation} 
in the norm topology of $\cB$, for {\it all} sequences $\alpha_n$ of
pairwise disjoint members of $\fS$.  The sum $\sum_{n=1}^\infty
\bnuv(\alpha_n)$ must therefore converge unconditionally in the norm.

The Banach space of interest to us is the histories Hilbert space
$\cH$ of \cite{hilbert} and we briefly review this construction
below. Let $V$ be the space of complex valued functions on $\fA$ which
are non-zero only on a finite number of elements of $\fA$. $V$ is the
free vector space over $\fA$ and the decoherence functional provides  an
inner product  
\begin{equation} \label{decoinner}
\langle u,v \rangle_V \equiv \sum_{\alpha \in \fA} \sum_{\beta \in \fA}
u^\ast (\alpha) v(\beta) D(\alpha, \beta).
\end{equation} 
$V$ is itself not a Hilbert space since it contains zero-norm vectors and
may not be complete. The histories Hilbert space $\cH$ is constructed
by taking the set of Cauchy sequences $\{ u_i\}$ in $V$ and
quotienting by the equivalence relation
\begin{equation} \label{equiv} 
\{u_i\} \sim \{v_i\} \quad \mathrm{if} \quad \lim_{i \rightarrow
  \infty} \parallel u_i - v_i \parallel_V =0,  
\end{equation} 
where the norm is given by the inner product.   We have 
\begin{eqnarray} 
[\{u_i\}] + [\{ v_i\}] &\equiv&  [\{ u_i+v_i\}] \nonumber \\ 
\lambda [\{ u_i\}] &\equiv& [\{\lambda u_i \}], 
\nonumber \\ 
\langle [\{u_i \}], [\{v_i\}] \rangle &\equiv & \lim_{i \rightarrow
  \infty} \langle u_i, v_i \rangle_V,    
\end{eqnarray} 
for all $[\{u_i\}], [\{ v_i\}] \in \cH, \, \, \lambda \in \bC$.

We define the {\sl quantum vector pre-measure} $\muv:\fA \rightarrow
\cH$ to be 
\begin{equation} 
\muv(\alpha) \equiv  [\chi_\alpha] \in \cH. 
\end{equation} 
where $\lbrack \cdot \rbrack$ denotes the
equivalence class under (\ref{equiv}) and we use the shorthand $\chi_\alpha$ to denote the constant Cauchy
sequence $\{ \chi_\alpha\}$ for the 
indicator function $\chi_\alpha : \fA \to \{0,1\}$ 
\begin{equation} 
\chi_\alpha(\beta) =  \left\{
              \begin{array}{ll}
                   1 & \textrm{ if $\beta=\alpha$},\\
                   0 & \textrm{ if $\beta\neq \alpha$}.
              \end{array}
       \right. 
\end{equation} 
Thus, 
\begin{equation}
\langle \muv(\alpha), \muv(\beta)  \rangle = D(\alpha, \beta), 
\end{equation} 
with the inner product taken in $\cH$. 
The bi-additivity of $D$ means that for any disjoint pair, $\alpha,
\beta \in \fA$ we have
\begin{equation} 
\parallel \chi_{\alpha\cup \beta}  - \chi_\alpha -\chi_\beta \parallel = 0
\end{equation} 
in the norm derived from the inner product \eqref{decoinner}. This
ensures that $\muv(\alpha \cup \beta) =\muv(\alpha) + \muv(\beta)$
and therefore  that $\muv$ is finitely additive: 
\begin{equation}
\muv\left(\bigcup_{i=1}^n \alpha_i\right)= \sum_{i=1}^n \muv(\alpha_i),    
\end{equation} 
for $n$ mutually disjoint sets $\alpha_i \in
\fA$. Hence $\muv$ is a vector pre-measure. 

We pause here to note that the term ``quantum vector (pre-)measure'' does not imply that $\muv$
fails to satisfy the Kolmogorov sum rule -- $\muv$ is a vector pre-measure and {\sl{is}} finitely
additive.  We use the phrase ``quantum vector (pre-)measure'' to emphasise the physical origins of
$\muv$ and that it is valued in the histories Hilbert space $\cH$ constructed from the quantum
measure.

As discussed in the introduction, an important question in measure theory is whether a pre-measure
defined on an algebra $\fA$ {\sl extends} to a measure on the sigma algebra $\fS_\fA$ generated by
$\fA$.  Specifically a vector measure $\bnuv: \fS_\fA \rightarrow \cB$ is said to be an {\sl
  extension} of a vector pre-measure $\nuv: \fA \rightarrow \cB$, if its restriction to $\fA \subset
\fS$ is $\bnuv|_\fA = \nuv$.  A vector pre-measure $\nuv$ on $\fA$ is then said to extend to $\fSA$
if there exists a unique vector measure $\bnuv:\fSA \rightarrow \cB$ such that $\bnuv|_\fA = \nuv$.

The extension $\bmuv$ of a quantum vector pre-measure $\muv$, if it exists, can be used to define a
decoherence functional $\bD: \fS_\fA \times \fS_\fA \rightarrow \bC$
\begin{equation}
\bD(\alpha, \beta) \equiv \langle \bmuv(\alpha), \bmuv(\beta) \rangle
\quad \forall \alpha, \beta \in \fS_\fA.    
\end{equation} 
The restriction $\bD|_\fA=D$, the decoherence functional on
$\fA$, and hence $\bD$ can be viewed as an 
extension of $D$ on $ \fA \times \fA$ to a functional on $\fSA \times
\fSA$.  $\bD$ is Hermitian, since 
\begin{equation} 
\langle \bmuv(\alpha), \bmuv(\beta)\rangle = \langle \bmuv(\beta),
\bmuv(\alpha)\rangle^\ast.  
\end{equation} 
Moreover, the countable additivity of $\bmuv$ and the countable biadditivity of the inner product
implies countable biadditivity of $\bD$. In addition, $\bD$ is strongly positive.  Finally, since
$\Omega \in \fA$, $\bD(\Omega, \Omega)=D(\Omega,\Omega)=1$. Thus, the extension $\muv$ can be used
to construct a countably biadditive, positive and Hermitian decoherence functional on $\fS_\fA$. If
$\bmuv$ is the unique extension of $\muv$, then so is $\bD$ the unique extension of $D$. (On the
other hand, a unique extension $\bD$ of $D$ yields a $\bmuv$ determined only up to an overall phase
which itself can be determined if $\muv$ is also known.)

For a non-negative scalar pre-measure $\mu:\fA \rightarrow \re^+$, the Carath\'eodory extension
theorem \cite{kac,halmos} guarantees the existence of a unique extension.  For complex vector
measures the Carath\'eodory-Hahn-Kluvanek extension theorem \cite{diesuhl} gives necessary and
sufficient conditions for a vector pre-measure on $\fA$ to extend to a vector measure on
$\fS_\fA$. In the case of finite dimensional vector measures, the extension question is equivalent
to the simpler question of the extension of the complex scalar component measures. Since our present
interest is in examining finite dimensional systems, we will not discuss the
Carath\'eodory-Hahn-Kluvanek theorem here, though it may be relevant to the larger program.  For our
purposes it suffices to focus on the property of {\sl bounded variation}. The {\sl total variation}
$\tmuv$ of a vector pre-measure $\muv$ is defined to be
\begin{equation}
\tmuv (\alpha)= \sup_{\pi(\alpha)} \sum_\rho \parallel
 \muv(\alpha_\rho) \parallel, 
\end{equation} 
where the supremum is over all finite partitions $\pi(\alpha)=\{
\alpha_\rho\}$ of $\alpha$ (note that $\tmuv(\alpha)$ is not just $|\muv(\alpha)|$). $\tmuv$ is itself a non-negative finitely
additive pre-measure on $\fA$ and is countably additive iff $\muv$ is
(Prop. 9, Chapter 1.1, \cite{diesuhl}).  $\muv$ is said to be of
{\sl bounded variation} if $\tmuv(\alpha) <\infty$ for all $\alpha \in
\fA$.

We note that in any basis the components $\muv^{(i)}$,
$i=1, \ldots, n$ of a vector pre-measure $\muv: \fS \rightarrow \bC^n$
are themselves complex-valued pre-measures on $\fS$.

\begin{claim} \label{components} 
Let $\muv: \fA \rightarrow \bC^n$ be a vector pre-measure and
${\muv}^{(i)}: \fA \rightarrow \bC$,  $i=1, \ldots n$ be the components of
$\muv$ in an orthonormal basis. Then $\muv$ is of bounded variation iff
${\muv}^{(i)}$ is of bounded variation. 
\end{claim} 
{\bf Proof:}  Since $\parallel \muv(\alpha) \parallel \, \geq \, \parallel
{\muv}^{(i)}(\alpha) \parallel$ for every $i \in \{1, \ldots, n\}$, 
\begin{equation}
\tmuv(\alpha) \geq |{\muv}^{(i)}|(\alpha) 
\end{equation} 
for each $i$. Therefore if $\muv$ is of bounded variation then so is
${\muv}^{(i)}$. From the triangle inequality 
\begin{equation}
 \parallel \muv(\alpha) \parallel \leq \sum_{i=1}^n |{\muv}^{(i)}(\alpha)|
\end{equation} 
Thus, for any finite partition $\{ \alpha_\rho \}$ of $\alpha$, $\rho \in
\{1, \ldots, m<\infty \} $
\begin{equation}
\sum_{\rho=1}^m  \parallel \muv(\alpha_\rho) \parallel  \leq \sum_{\rho=1}^m
\sum_{i=1}^n |{\muv}^{(i)}(\alpha_\rho)| \leq \sum_{i=1}^n |{\muv}^{(i)}(\alpha)|, 
\end{equation} 
Clearly, if $|{\muv}^{(i)}|(\alpha) \leq b_i < \infty$ , i.e.,
it is bounded for each $i$, then $$\sum_{\rho=1}^m \parallel\muv(\alpha_\rho) \parallel < \sum_i b_i $$ for every finite partition of
$\alpha$. Hence $\muv$ is also bounded.  \hfill $\qed$ \vskip0.3cm 

For a complex measure countable additivity implies that its total variation is bounded
\cite{rudin}. The components $\bmuv^{(i)}$ $i=1, \ldots , n$ of a countably additive vector measure
$\bmuv:\fS \rightarrow \bC^n$ are also countably additive, and hence are of bounded variation. Thus,
by the above Claim, the countable additivity of $\bmuv$ implies that it is of bounded variation.

Bounded variation of a complex measure implies its restriction $\mu|_\fA$ to any subalgebra $\fA
\subset \fS$ is also of bounded variation. Thus, a necessary condition for a complex pre-measure on
$\fA$ to extend to a measure on $\fSA$ is that it is of bounded variation.  Along with the above
results this means that

\begin{claim}
  Bounded variation is a necessary condition for a finite dimensional vector pre-measure $\muv: \fA
  \rightarrow \bC^N$ to extend to a vector measure $\bmuv: \fSA \rightarrow \bC^N$.
\end{claim}

\section{Finite Unitary Systems} \label{finite}

In this section we consider the class of finite $N$ dimensional
systems which evolve unitarily in discrete unit time steps, $t =1,2,3\dots$.
In the standard Hilbert space formulation, the Hilbert space at time
$t=m$ is $\cH_m = \bC^N$. The evolution of a state $ \psi \in \cH_1$
at an initial time $t=1$ to a state $\psi_m \in \cH_m$ at time $t=m$
is governed by the $N\times N $ single-step unitary matrices $U(k+1,k)$:
\begin{equation}
\psi_m = U(m,m-1)U(m-1,m-2)\dots U(2,1) \psi\,.
\end{equation}
Let $\{e_1,e_2,\dots e_N\}$ be an orthonormal basis for 
$\cH_1$.

To describe this system as a quantum measure space we first identify a history as an infinite string
$\ginf = (s_1, s_2, \ldots, s_i, \ldots )$, where each entry $s_i \in \{ 1, \ldots, N \}$.  The
configuration space has $N$ sites, each associated with one of the basis vectors $e_s$. $\Omega$ is
the infinite collection of all such strings and the event algebra $\fA$ is generated as follows.  We
associate with every length $m$ finite string $\gamma = ( s_1, s_2\ldots s_m)$ a {\sl cylinder set}
\begin{equation} \label{defcyl} 
\cyl(\gamma) \equiv \{\ginf  \in \Omega| \ginf(i)=s_i, \, \,  i=1,
2, \ldots m \}\,,  
\end{equation}
which is the set of histories for which the first $m$ entries are
specified by the string $\gamma$ but are unspecified thereafter.  If
$\gamma=(s_1,s_2, \ldots, s_m)$ and $\gamma'=(s_1',s_2', \ldots,
s_m')$, with $m'> m$, then  
\begin{eqnarray} \label{cylsubset} 
\cyl(\gamma') \subset \cyl(\gamma) \, && \mathrm{if}   \,\, \, s_i =s_i'
\, \, \forall \, \,  i \in \{ 1, 2, \ldots m\} \\  
\cyl(\gamma') \cap \cyl(\gamma) =\emptyset && \mathrm{otherwise}.   \label{cylint} 
\end{eqnarray} 
If $\fA$ represents the algebra generated from finite unions and intersections of these cylinder
sets, then (\ref{cylsubset}) and (\ref{cylint}) imply that any $\alpha \in \fA$ can be expressed as
a finite disjoint union of cylinder sets. In particular, for every $\alpha \in \fA$ there exists an
$m$ and a $k \leq N^m$ such that $\alpha = \cup_{i=1}^k \cyl({\gamma}_k)$ where the $\gamma_k$ are
strings of length $m$ and the $\cyl({\gamma}_k)$ are mutually disjoint. Here, $N^m$ is the number of
possible length $m$ strings.

The decoherence functional for such a unitary system is given by
\begin{equation}\label{unitary} 
D(\cyl(\gamma), \cyl(\gamma'))= A^*(\gamma) A(\gamma')
\delta_{s_ms_m'},    
\end{equation} 
where $\gamma, \gamma'$ are length $m$ finite strings (or ``truncated
histories'') and $A(\gamma_i)$ is a complex amplitude. Assuming an
initial  state $\psi \in \cH_1$ 
\begin{eqnarray}  
\label{df}
 A(\cyl(\gamma)) & = & U_{s_m s_{m-1}}(m,m-1) \ldots 
U_{s_3 s_2}(3,2) U_{s_2 s_1}(2,1) \psi(s_1).  \nonumber \\
\end{eqnarray}
Here $U_{s_is_{i-1}}(i,i-1)$ is the amplitude to go from $s_{i-1}$ at
$t=i-1$ to $s_i$ at $t=i$ and $\psi(s_1) = <e_{s_1}, \psi>$.
The decoherence functional on the full event algebra
$\fA$ is then obtained using its bi-additivity property.

The {\sl restricted evolution} from the initial state $\psi \in
\cH_1$ with respect to a truncated history $\gamma $ is defined to be 
\begin{equation} \label{restricted}  
 {\tpsi_\gamma}  \equiv \hC_{\gamma} \psi   \, \, \in
  \cH_{m},  
\end{equation}
where the {\sl class operator} 
\begin{equation}\label{classop}  
\hC_{\gamma} \equiv  P_{s_m}U(m,{m-1})P_{s_{m-1}}U(m-1,m-2)\ldots
U(2,1) P_{s_1}
\end{equation}
and $P_s$ is the projector that projects onto the basis
state $e_s$.  
Evolving $ {\tpsi_\gamma}  $ back to the initial time gives us the
state 
\begin{equation} 
 \psi_\gamma  \equiv (U(m,m-1)U(m-1,m-2)\dots U(2,1))^\dagger  \hC_{\gamma}  \psi  \, \,
 \in \cH_1\,,
\end{equation}  
and it can be shown that the decoherence functional 
\begin{equation} 
D(\cyl(\gamma), \cyl(\gamma')) \equiv \langle \psi_\gamma ,
\psi_{\gamma'}  \rangle \,.
\end{equation} 
  For any $\alpha \in \fA$ and a partition 
 into cylinder sets $\alpha=\cup_{i=1}^k \cyl({\gamma}_k)$,
one can also define the state
\begin{equation}
  \psi_\alpha   = \sum_{k}  \psi_{{\gamma}_k}         
\end{equation} 
and show that 
\begin{equation}
  \langle \psi_\alpha,   \psi_\beta  \rangle  =  D(\alpha, \beta).  
\end{equation}

In \cite{hilbert} it was shown that for a finite $N$ dimensional system which evolves unitarily with
discrete time steps there is, generically, an explicit, physically meaningful isomorphism $f: \cH
\rightarrow \cH_1$ between the histories Hilbert space $\cH$ and the standard Hilbert space
$\cH_1=\bC^N $. Namely,
\begin{equation}
  f([\{ u_i \} ])  \equiv \lim_{i \rightarrow \infty}   f_0(u_i), 
\end{equation}  
with $f_0: V \rightarrow \cH_0$ given by
\begin{equation} 
 f_0(u)  \equiv \sum_{\alpha \in \fA }  u(\alpha)  \psi_\alpha. 
\end{equation}  
Here $\langle f_0(u), f_0(v)  \rangle  = \langle u, v  \rangle$  and
hence  $\langle f([\{ u_i \}]), f([\{ v_i\}])\rangle   = \langle [\{
u_i \}], [\{ v_i\}] \rangle $. 
The quantum
vector pre-measure $\muv: \fA \rightarrow \cH$ is therefore mapped to a
vector measure $\wmuv\equiv f\circ \muv$ on $\cH_1$ 
\begin{eqnarray} 
 \wmuv(\alpha)   & = &  \sum_{\beta \in \fA} \chi_\alpha(\beta)   
 \psi_\beta     \nonumber \\  &=&     \psi_\alpha    .  
\end{eqnarray}

\begin{claim} 
The quantum vector measure for a generic (to be defined) finite unitary
system with discrete time steps is not of bounded variation. 
\end{claim} 

{\bf Proof:} In what follows we will first assume that $U$ is time
independent so that $U(k+1,k)=U$ for all $k$, where $U$ is the single time-step evolution
operator. 

For a string of length $m$, $\gamma=(j_1, j_2, \ldots
j_m)$
\begin{equation}
\wmuv(\cyl(\gamma))=(U^\dagger)^m U_{j_mj_{m-1}} U_{j_{m-1} j_{m-2}}
\ldots U_{j_2j_1} \beta_{j_1}  e_{j_1} 
\end{equation}  
where $U_{jk}=\langle e_j,   U  e_k \rangle $ and $\beta_{j_1}= \langle
e_{j_1},    \psi  \rangle $, so that
\begin{equation}
 \parallel \wmuv(\cyl(\gamma)) \parallel =|\beta_{j_1}|
 |U_{j_mj_{m-1}} U_{j_{m-1} j_{m-2}} \ldots U_{j_2j_1}|.  
\end{equation} 
A string $\gamma'$ of length $m+k$ will be said to be an extension of
$\gamma$ if the first $m$ entries of $\gamma'$ are the same, so that
$\cyl(\gamma') \subset \cyl(\gamma)$. The set of all $n+k$ extensions
$\{\cyl(\gamma^{(i_1i_2\ldots i_k)}) \}$ of $\gamma$ thus provides a
partition of $\cyl(\gamma)$ so that 
\begin{eqnarray} \label{tot}  
\twmuv(\cyl(\gamma))  & \geq &  \sum_{i_1,i_2 \ldots i_k=1}^N  \parallel
\wmuv(\cyl(\gamma^{(i_1i_2\ldots i_k)}) \parallel \nonumber \\ 
&= & \biggl( \sum_{i_1,i_2 \ldots, i_k=1}^N |U_{i_k
  i_{k-1}}||U_{i_{k-1}i_{k-2}}| \ldots |U_{i_1 j_m}|\biggr)
\nonumber \\     
&& \qquad \times \parallel 
\wmuv(\cyl(\gamma)) \parallel \,.
\end{eqnarray}

Since $U$ is unitary, $\sum_{i=1}^N |U_{ij}|^2=1$ and $\sum_{j=1}^N
|U_{ij}|^2=1$. Thus, 
\begin{eqnarray} 
\sum_{i=1}^N |U_{ij}| &= &  1+\zeta_j  \\ 
\sum_{j=1}^N |U_{ij}| &= & 1+\eta_i
\end{eqnarray} 
where $\zeta_j \geq 0$ and $\eta_i \geq 0$. If $\zeta_j=0$ then there exists an $i$ such that
$|U_{ij}|=1$ and $U_{i'j}=0$ for all $i'\neq i$. Similarly, for every $j'\neq j $,
$U_{ij'}=0$. Hence the $i$th row and the $j$th column each have a single non-zero entry $U_{ij}$
which is pure phase.

We now define the genericity assumption for the time-independent case to be that $U$ does not have
such entries, i.e., $\zeta_j > 0$ for all $j$, and hence $\eta_i>0$ for all $i$.  This excludes
dynamics which consist of simple permutations of a number of the site. This assumption means that
there is a smallest strictly positive $\zeta \leq \zeta_j$ for all $j$. This allows us to
iteratively bound the term in brackets in (\ref{tot})
\begin{eqnarray} \label{factoring}
\sum_{i_{k-1}, \ldots i_1 =1}^N \biggl( \sum_{i_k} |U_{i_k
  i_{k-1}}| \biggr)&\times&\!\! |U_{i_{k-1}i_{k-2}}| \ldots |U_{i_1j_m}|
  \nonumber  \\  
&=& \!\!\!
\sum_{i_{k-2}, \ldots i_1 =1}^N \biggl(\sum_{i_{k-1}=1}^N (1+
  \zeta_{i_{k-1}}) |U_{i_{k-1}i_{k-2}}|\biggr) \ldots
 |U_{i_1j_m}|  \nonumber \\ 
& \geq & (1+\zeta) \sum_{i_{k-2}, \ldots i_1 =1}^N  \biggl( 
  \sum_{i_{k-1}=1}^N  |U_{i_{k-1}i_{k-2}}|\biggr) \ldots
  |U_{i_1j_m}|  \nonumber \\ 
& \geq & (1+\zeta)^{k}. 
\end{eqnarray} 
Hence for every  $k$, 
\begin{equation} \label{zbv} 
\twmuv(\cyl(\gamma)) \geq (1+\zeta)^{k} \times \parallel
\wmuv(\cyl(\gamma)) \parallel     
\end{equation}  
Since  $\zeta >0$, $\twmuv$ is not bounded.

For time dependent $U$ the argument is similar,  except
that the $\zeta$ are now time dependent. In particular, 
instead of extracting a time-independent factor $(1+\zeta)$ 
in Eqn (\ref{factoring}) at every step,  it becomes time-dependent so
that 
\begin{equation} \twmuv(\cyl(\gamma)) \geq \prod_{r=1}^{m} (1+\zeta_r) \parallel
\wmuv(\cyl(\gamma)) \parallel. \label{lasteq}
\end{equation} 
where $\sum_{i=1}^N |U_{ij}(r,r-1)| = 1+\zeta_j(r)$ and $\zeta_r$ is the lowest value of
$\zeta_j(r)$ as one varies over $j$. Again, assuming that $\zeta_r>0$, the product $ \prod_{r=1}^{m}
(1+\zeta_r)$ converges as $m\rightarrow \infty$ only if $\sum_{r=1}^m \zeta_r$ does.  The genericity
condition is therefore that $\sum_{r=1}^m \zeta_r$ diverges as $m \rightarrow \infty$ in which case
(\ref{lasteq}) diverges and the pre-measure is not of bounded variation.  \hfill $\qed$ \vskip0.3cm
\vskip 0.2cm


Thus the quantum vector pre-measure $\wmuv$ on $\fA$ does not
extend to a quantum vector measure on $\fS_\fA$ for generic finite
dimensional unitary systems with discrete time steps.

\section{Complex Percolation} \label{cp} 

In causal set theory, the histories space of continuum spacetime geometries is replaced by the
collection of causal sets.  A causal set $C$, as defined in \cite{blms}, is a locally finite
partially ordered set, namely, a countable collection of elements, with an order relation $\prec$
which for all $x,y,z \in C$ is (i) transitive ($x \prec y\, , y \prec z \, \Rightarrow x \prec z$),
(ii) {\sl irreflexive}, ($x \not\prec x$) and (iii) locally finite i.e. if $Past(x) \equiv \{w \in
C| w \prec x \}$ and $Fut(x)\equiv \{w \in C| w \succ x \} $ then the cardinality of the set
$Past(x) \cap Fut(y)$ is finite.  We say that two elements are {\sl {linked}} if they are related in
the order but there is no element between them in the order.

A causal set is a model for discrete spacetime in which the elements of $C$ represent spacetime
events and the partial order represents the causal relationships between events. A generic causal
set has no continuum spacetime approximation, however, and it is only via an appropriate choice of
dynamics on the set of all possible causal sets that one expects a continuum spacetime to emerge.

The transitive percolation dynamics for causal sets is a classical stochastic dynamics and was
studied in detail in \cite{csg,davethesis}, and is determined by a single coupling constant $p\in
[0,1] $. Here, a causal set is ``grown'' element by element starting with a single element.  At
stage $n>1$ the $n$th element $e_n$ is born and, for each $k =1,2\dots n-1$ independently, $e_n$ is
put to the future of $e_k$ with probability $p$ or with probability $1-p$ left unrelated to $e_k$.
The transitive closure is then taken and the stage $n$ is complete. The resulting causal set grown
in this way is ``labelled'' by the growth, with each element labelled by the stage at which it is
born (see Fig \ref{postcau}).  Thus, the growth process is stochastic and produces a labelled causal
set of infinite cardinality in the asymptotic limit $n \rightarrow \infty$. Such a causal set is
always {\sl past finite}, i.e., the cardinality of the past set $Past(x)=\{y| y\prec x \}$ is finite
for all $x \in C$. Even though the causal sets produced are labelled, the resultant dynamics
satisfies a discrete form of general covariance in that the probabilities of growing, by stage $n$,
two labelled causal sets are the same if there is an order preserving isomorphism between them. In
addition, the dynamics satisfies the ``Bell causality'' condition described in \cite{csg}.  The
labelling of a causal set produced via a growth is always {\sl order preserving} namely, for any $e,
e' \in C^n$, $e \prec e'$ implies that $l(e) < l(e')$, where $l(e)$ is the label of the element $e$.

Figure \ref{postcau} show the first few stages of this growth process. Let $\cn_c$ denote the {\sl
  $n$-chain} or the totally ordered set of $n$ elements and $\cn_a$ the {\sl $n$-antichain} or the
set of $n$ mutually unrelated elements. Starting with the first element $e_1$ the second element
$e_2$ is added with probability $p \in [0,1]$ to the future of $e_1$, to get the (uniquely) labelled
$2$-chain $C^2_c$ or is left unrelated to $e_1$ with probability $q$ to get the (uniquely) labelled
$2$-antichain $C^2_a$.  Since for $n=2$ these are the only possible $2$ element labelled causal
sets, $p+q=1$ (see Fig \ref{postcau}). Subsequently, the three $3$-element causal sets are grown
from $C^2_c$ and $C^2_a$ as shown in Fig \ref{postcau}.  In the first transition from $C^2_c$ in the
figure (from the left), $e_3$ is added to the immediate future of $e_2$ with probability $p$, in the
second, $e_3$ is added to the immediate future of $e_1$ but is unrelated to $e_2$ with probability
$p \times q$, and in the third, $e_3$ is unrelated to both $e_1$ and $e_2$ with probability $q^2$.
In the figure we see that the middle three $3$-element labelled causal sets are different order
preserving relabellings of the same unlabelled causal set.

\begin{figure}[ht]
\centering \resizebox{3.75in}{!}{\includegraphics{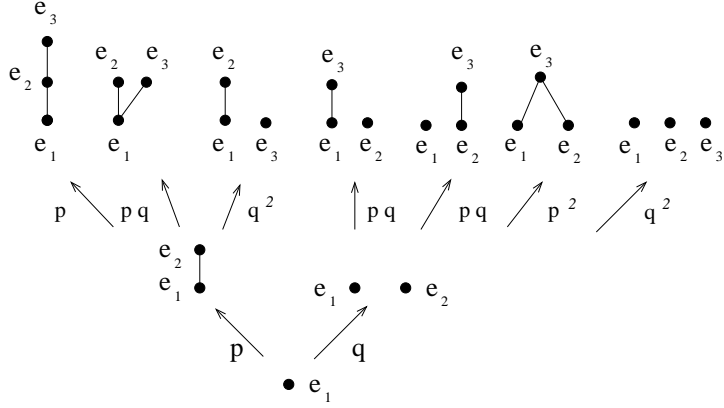}}
\vspace{0.5cm}
\caption{{\small Transition probabilities up to 3-element labelled causal sets.}}\label{postcau} 
\end{figure}

The probability $P(C^{n})$ for an $n$-element labelled causal set $\cn$ is equal to the product of
the transition probabilities. As can be verified from the examples in Fig \ref{postcau}, this
probability is independent of the labelling: in general, the probability for a labelled causal set
is given by $p^L q^{\binom{n}{2}-R}$ where $L$ is the number of links and $R$ the number of
relations \cite{csg}.  The transition probability for going from an $n$-element causal set $C^n$ --
a {\sl {parent}} -- to one of its {\sl {children}} $C^{n+1}$ is given by a product of $p$'s and
$q$'s: if the new element is to the immediate future of ({\sl {i.e.}} linked to) $\fu$ elements in
$\cn$ and unrelated to $\un$ elements, then the transition amplitude is $p^\fu q^\un$. For a given
parent $\cn$ therefore, one can assign an index set $\iS(\cn)$ of all pairs $(\fu,\un)$, some of
which may repeat.  For example, when the parent is $C^2_c$, the index set is $\iS(C^2_c)= \{(1,0),
(1,1), (0,2) \}$ while if the parent is $C^2_a$, the index set is $\iS(C^2_a)= \{(1,0), (1,0),
(2,0),(0,2) \}$ (see Fig \ref{postcau}). Note that the set $\iS(\cn)$ can contain repeated
entries.

It is useful to point to two special types of transitions which will make their appearance in our
analysis below. The first is the ``timid transition'' in which the new element is added to the
future of all the elements in $\cn$, so that the transition probability is $p^m$ where $m$ is the
number of maximal elements in $\cn$. Thus, $(\fu,\un)= (m, 0)$. The second is the ``gregarious
transition'' in which the new element is unrelated to all the existing elements in $\cn$, so that
the transition probability is $q^n$ and $(\fu,\un)= (0, n)$.

Let $\Omega$ be the set of all infinite, past finite, labelled causal sets. If $\cn(k)$ refers to
the $k$th labelled element in $\cn$, the cylinder set associated with $\cn$ is
\begin{equation} 
\cyl(\cn)\equiv \{c \in  \Omega | c(k)=\cn(k), \, k=1, \ldots,  n \} 
\end{equation} 
where $c(k)$ is the $k$th element of $c$.  The event algebra $\fA$ is the set of all finite unions
of these cylinder sets and it can be shown that every element of $\fA$ is equal to a finite union of
mutually disjoint cylinder sets.  In particular $\cyl(\cn)$ is equal to the union of the (disjoint)
cylinder sets of all the children of $\cn$: $ \cyl(\cn)= \bigcup_i \cyl(C^{n+1}_i)$. However, unlike
the finite unitary systems, the number of $n+1$-element children $\{ C^{n+1}_i\}$ depends on the
parent $\cn$.

Complex percolation is a natural quantum generalisation of transitive percolation in which real
probabilities are replaced by complex amplitudes. Thus, the real parameter $p$ of transitive
percolation is made complex and gives the transition amplitude for the newly born element to be put
to the future of each existing element, while $q$ is the transition amplitude for it to be unrelated
(we will see later that $p+q =1$).  The decoherence functional for complex percolation has a simple
product form
\begin{equation}\label{product} 
D(\cyl(C^n),\cyl({C'}^n))=A^*(C^n) A({C'}^n) 
\end{equation} 
for cylinder sets, where $A(C^n)$ is the amplitude for the transition from the empty set to the
$n$-element causal set $C^n$.  Finite biadditivity of $D$ and finite additivity of $A$ are
equivalent and $A$ is a complex measure on $\fA$.  The normalisation condition $D(\Omega, \Omega)=1$
implies that $|A(\Omega)|=1$. We have for $\alpha, \beta \in \fA$,
\begin{equation} \label{product} 
D(\alpha, \beta)= A^*(\alpha) A(\beta).
\end{equation}
It is easy to demonstrate that for any such ``product'' decoherence functional the histories Hilbert
space $\cH \simeq \bC$.  Choose a vector $v \in V$, with $\parallel v \parallel \neq 0$.  We show
that for every $u \in V$ there exists a $\lambda \in \bC$ such that $\{ u \} \sim \lambda \{ v\}$,
where $\sim$ is the equivalence relation Eqn (\ref{equiv}) and $\{u \}, \{v\} $ are the constant
Cauchy sequences for $u$ and $v$ respectively. Then
\begin{equation} 
\parallel u-\lambda v \parallel^2 = |S_1|^2 -(\lambda S_1^*S_2 + \lambda^* S_1
S_2^*) + |\lambda|^2 |S_2|^2 = |S_1 - \lambda S_2|^2,   
\end{equation} 
where $S_1= \sum_{\alpha \in \fA} A(\alpha)u(\alpha)$ and $S_2=
\sum_{\alpha \in \fA} A(\alpha)v(\alpha)$. This factorisation is
possible because of the product form Eqn (\ref{product}). If we then
choose $\lambda=S_1/S_2$ ($S_2 \neq 0$) we have $\parallel u-\lambda
v \parallel =0$.

In particular, since for any $\alpha \in \fA$ and a finite partition $\pi(\alpha)=\{\alpha_\rho \}$
of $\alpha$
\begin{equation} 
\tmuv(\alpha) \geq \sum_\rho \parallel \muv(\alpha_\rho) \parallel = \sum_\rho
|A(\alpha_\rho)|  
\end{equation}  
Hence $\tmuv=|A|$, where $|A|$ the total variation of $A$, and
$\muv$ is equal to $A$ up to a phase.  

For the special case $p$ real and $p \in [0,1] $, the transition amplitudes are the same as the
transition probabilities of classical transitive percolation.  However, this {\sl real quantum
  percolation} model is {\it distinct} from transitive percolation since the quantum measure
$D(\alpha, \alpha)= |A(\alpha)|^2$ and is therefore non-additive. However, the amplitude measure $A$
is additive and non-negative, and the Carath\'eodory-Kolmogorov extension theorem implies that the
quantum vector pre-measure extends to the full-sigma algebra.  As we will presently see, this
special case is the only one which does admit such an extension.

We first show that $q$ must equal $1-p$ . The normalisation condition on $D$ means that
$|A(\Omega)|=1$, so that $A(\Omega)=\exp(i\Phi)$ and we will choose this phase to be 1. If $C^1$
denotes the single element causal set then $\cyl(C^1)=\Omega$. We also have $\cyl(C^1)= \cyl(C^2_c)
\cup \cyl(C^2_a)$ and $\cyl(C^2_c) \cap \cyl(C^2_a) = \varnothing$, where, as before $C^2_c$ and
$C^2_a$ are the 2-chain and 2-antichain respectively.  This means
\begin{eqnarray} \label{Phi}
A(\cyl(C^1)) & = & A(\cyl(C_c))+A(\cyl(C_a)) \nonumber \\ 
&=& A(\cyl(C^1)) \times (p+q) \, \,  
\Rightarrow \, \, p+q=1. \nonumber 
\end{eqnarray}  
As shown above, the histories Hilbert space for a
product decoherence functional is 1-dimensional. This means that the
quantum vector measure is, up to an overall phase, just the amplitude
$A:\fA \rightarrow \bC$.

We now show that
\begin{lemma} 
The quantum vector measure of complex percolation is not of bounded
variation when the parameter $p$ is not real.
\end{lemma}

We begin by considering the set of all labelled $n$-element causets $\{ \cn_i \}$, where $i=1,2,
\ldots I(n)$ where $I(n)$ is the number of $n$ element labelled causal sets. For example, for $n=2$,
$I(2)=2$ so that $i=1,2$ and for $n=3$, $I(3)=7$, so that $i=1, \ldots, 7 $.  Since $A(\Omega) =
\sum_{i=1}^{I(n)} A(\cyl(\cn_i))$, and $|A(\Omega)|=1$, the triangle inequality implies that
\begin{equation} \label{triangle} 
\sum_{i=1}^{I(n)} |A(\cn_i)| \geq 1,   
\end{equation} 
where for brevity of notation we have replaced $\cyl(\cn)$ with $\cn$. The equality is satisfied
only if all the $A(\cn_i)$ are collinear. We see that

\begin{figure}
\begin{center}
\setlength{\unitlength}{0.8cm}
\begin{picture}(5,6.5)(-3,0.5)

\put(-3,1){\circle*{.1}}
\put(-3,2){\circle*{.1}}
\put(-3.02,2.8){\vdots}
\put(-3,4){\circle*{.1}}
\put(-3,5){\circle*{.1}}
\put(-3,6){\circle*{.1}}

\put(-3,1){\line(0,1){1.5}}
\put(-3,6){\line(0,-1){2.5}}

\put(-2.8,1){$e_1$}	
\put(-2.8,2){$e_2$}	
\put(-2.8,4){$e_{n-2}$}	
\put(-2.8,5){$e_{n-1}$}	
\put(-2.8,6){$e_{n}$}	


\put(1,1){\circle*{.1}}
\put(1,2){\circle*{.1}}
\put(0.98,2.8){\vdots}
\put(1,4){\circle*{.1}}
\put(1,5){\circle*{.1}}
\put(0.5,6){\circle*{.1}}
\put(1.5,6){\circle*{.1}}

\put(1,1){\line(0,1){1.5}}
\put(1,5){\line(0,-1){1.5}}
\put(1,5){\line(1,2){0.5}}
\put(1,5){\line(-1,2){0.5}}

\put(1.2,1){$e_1$}
\put(1.2,2){$e_2$}
\put(1.2,4){$e_{n-3}$}
\put(1.2,5){$e_{n-2}$}
\put(0.1,6.2){$e_{n-1}$}
\put(1.4,6.2){$e_{n}$}

\end{picture}
\caption{Hasse diagrams for $\cn_c$ and $\cn_v$.}
\label{HasseD}
\end{center}
\end{figure}
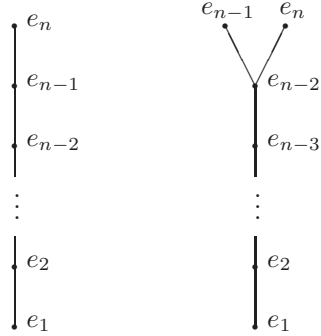

\begin{claim} For any $n\geq 2$, the equality in (\ref{triangle}) is
  satisfied only when the parameter $p$ is real. 
\end{claim} 
\noindent {\bf Proof:} Let $p=|p|\exp^{i\theta}$, $q=|q|\exp^{i \phi}$.  First, for $n=2$ the
equality means that $|p| + |q| = 1 $, which combined with $p+q=1$ means that $p$ is real and
non-negative.  For $n > 2$, consider the following two $n$ element causal sets, (a) the $n$ chain
$\cn_c$ and (b) $\cn_v$, an $n-2$ chain topped with a ``$V$'', i.e. with $e_n$, $e_{n-1}$ to the
immediate future of the maximal element $e_{n-2}$ of the $n-2$ chain $C^{n-2}_c$, and unrelated to
each other (see Fig \ref{HasseD}).  The amplitudes for these causal sets are $A(\cn_c)=p^{n-1}$, and
$A(\cn_v)=p^{n-1}q$.  Requiring collinearity of these amplitudes is therefore equivalent to
requiring that $p$ is real and non-negative.  \hfill $\qed$ \vskip0.3cm

Thus, for $p$ non-real 
\begin{equation} 
\sum_{i=1}^{I(n)} |A(\cn_i)| >  1.  
\end{equation} 
for all $n \geq 2$.  For $n=2$ it is useful to express the inequality as
\begin{equation} 
|p|+|q|=1+\zeta,  \, \,  \zeta>0
\end{equation} 

\begin{claim} If $|p|>1$ or $|q|>1$, the quantum vector measure is not
  of bounded variation. 
\end{claim} 

\noindent {\bf Proof:} Consider any partition of $\Omega$ which contains $\cyl(\cn_c)$. Since
$A(\cn_c)=p^{n-1}$, $|A|(\Omega) >|p|^{n-1}$ is a strict inequality for any $n$ if $q\neq 0$.
Similarly, consider any partition of $\Omega$ which contains $\cyl(\cn_a)$. Since
$A(\cn_a)=q^{n(n-1)}$, $|A|(\Omega) > |q|^{n(n-1)}$ for any $n$, if $p\neq 0$. Thus, $A$ is not of
bounded variation if either $|p|>1$ or $|q|>1$. \hfill $\qed$ \vskip0.3cm

Thus, we may restrict our attention to $|p| <1, |q| <1$. Consider an n-element causal set
$C^{n}$. Let $\{C^{n+1}_{j_1} \}$ be the set of its $n+1$ element descendants and $a(j_1)$ the
associated transition amplitude.  The index $j_1 = 1, 2, \ldots J_1(\cn)$ where $J_1(\cn)$ are the
number of descendants of $\cn$.  In turn, let $ C^{n+2}_{j_1j_2}$ denote the $n+2$ element
descendant of $C^{n+1}_{j_1}$ and $a_{j_1}(j_2)$ the associated transition amplitude, and so on.
The index $j_2$ depends on $j_1$ since $j_2 = 1, 2, \ldots J_2(\cn_{j_1})$ where $J_2(\cn_{j_1})$
are the number of descendants of $C^{n+1}_{j_1}$ and so on. $j_2$ thus carries a hidden index $j_1$,
but we will not include it explicitly in the expressions below.  The set $\Pi \equiv
\{C^{n+s}_{j_1j_2\ldots j_s} \} $ of $n+s$ element descendants of $C^n$ provides a disjoint
partition of $\cyl(C^n)$ where the range $J_r(C^{n+r-1}_{j_1j_2\ldots j_{r-1}})$ of each $j_r$ is
determined by its parent $C^{n+r-1}_{j_1j_2\ldots j_{r-1}}$. Thus the total variation
\begin{equation}
|A|(C^n) \geq \sum_{j_1=1}^{J_1} \left( \sum_{j_2=1}^{J_2(j_1)}\left(
 \ldots \left( \sum_{j_s=1}^{J_s(j_{s_1}j_{s-2} \ldots j_1)}
 |A(C^{n+s}_{j_1j_2\ldots j_s})|\right)\ldots  \right)\right)    
\end{equation} 
where 
\begin{equation} 
A(C^{n+s}_{j_1j_2\ldots j_s})=A(C^n)\times a(j_1)a_{j_1}(j_2) \ldots
a_{j_{s-1}}(j_s) 
\end{equation} 
Thus 
\begin{eqnarray} \label{nested} 
 |A|(C^n) \geq |A(C^n)| \times\left( \sum_{j_1} |a(j_1)| \left(\sum_{j_2} |a_{j_1}(j_2)| \left( \right. \right. \right.
\ldots \nonumber\\
\left. \left. \left. \left(\sum_{j_s} |a_{j_{s-1}}(j_s)| \right)\ldots \right)\right)\right), 
\end{eqnarray} 
where we have suppressed the dependencies of the $j_r$'s.  We now show that \begin{equation}
  \sum_{j_i}|a_{j_{i-1}}(j_{i})| \geq 1+ \zeta
\end{equation} 
for every  $i$. 
  
The final sum within the nested brackets of (\ref{nested}) 
\begin{equation} \label{sone} 
\sum_{j_s} |a_{j_{s-1}}(j_s)| \geq 1 
\end{equation} 
since 
\begin{equation}\label{stwo} 
\sum_{j_s} a_{j_{s-1}}(j_s)=1.  
\end{equation}  
Now, as in transitive percolation, each term in (\ref{stwo}) is of the form $p^\fu q^\un$, with
$(\fu,\un) \in \iS$, where we have suppressed the dependence of the index set $\iS$ on the parent
$C^{n+s-1}_{j_1j_2\ldots j_{s-1}}$. If $m$ is the maximal number of elements in
$C^{n+s-1}_{j_1j_2\ldots j_{s-1}}$, and $\iS'$ is the index set which excludes $(\fu, \un)=(m,0)$,
then
\begin{eqnarray} \label{unmod} 
\sum_{j_{s}} a_{j_{s-1}}(j_s) &= & \sum_{(\fu,\un) \in \iS}
  p^{\fu}q^{\un} \nonumber \\ 
&=& p^m + \sum_{(\fu,\un) \in \iS'}  p^{\fu}q^{\un}
\nonumber \\ 
 &= &  1+ \sum_{\ww \in W} \cw q^\ww,  
\end{eqnarray} 
for some appropriate index set $W$ and coefficients $\cw$. Since the above sum is always equal to
$1$, and is true for all $q$, this means that $\cw=0$. Thus, 
\begin{eqnarray} 
\sum_{j_s} |a_{j_{s-1}}(j_s)| &= & |p|^m+ \sum_{(\fu,\un)\in \iS'} |p|^{\fu}|q|^{\un} \nonumber \\
&=& (1+\zeta -|q|)^m + \sum_i(1+\zeta -|q|)^{u_i}|q|^{v_i} \nonumber
\\ 
&=& (1-|q|)^m  + \sum_{(\fu,\un)\in \iS'} (1-|q|)^{\fu}|q|^{\un} +  \nonumber \\ 
& & {m \choose 1} (1-|q|)^{m-1} \zeta + {m \choose 2}
  (1-|q|)^{m-2} \zeta^2 \ldots + \zeta^m + \nonumber \\ 
&& \sum_{(\fu,\un)\in \iS''}  \biggl( \quad {\fu \choose 1}
(1-|q|)^{\fu-1} \zeta + {\fu \choose 2} (1-|q|)^{\fu-2} \zeta^2 + 
\nonumber  \\ 
&& \qquad \ldots + \,\, \zeta^{\fu}  \quad \biggr)\,\, |q|^{\un},  
\label{expansion}
\end{eqnarray} 
where $\iS''$ is the index set which excludes both $(\fu, \un)=(m,0)$ and $(\fu,\un) =(0,n+s-1)$.
The first two terms are of the same form as expression (\ref{unmod}) and hence equal to $1$ since it
is independent of the choice of $|q|$.  Since $0\leq |q| \leq 1$, each of the terms in the above
expression is positive. It therefore suffices to focus on the terms linear in $\zeta$. We see that
this can be simplified to
\begin{eqnarray}  
&& \biggl(  (1-|q|)^{m} + \sum_{(\fu,\un)\in \iS''}
(1-|q|)^{\fu}|q|^{\un}
\biggr)(1-|q|)^{-1}  \nonumber \\ 
&&+ \biggl((m-1) (1-|q|)^{m-1} + \sum_{(\fu,\un)\in \iS''} \biggl( (\fu-1)
(1-|q|)^{\fu-1}|q|^{\un}\biggr)\biggr) \nonumber \\ 
&>& \biggl( 1 - |q|^{n+s-1})
\biggr)(1-|q|)^{-1}.  
\label{linear}
\end{eqnarray} 
For any $n+s-1 \geq 1$ the above expression is $> 1$ for $|q|<1$.  Thus, each nested sum $\sum_{j_s}
|a_{j_{s-1}}(j_s)| \geq (1+\zeta)$, for any $j_{s-1}$ and hence from (\ref{nested}) we see that
\begin{eqnarray} 
|A|(C^n) &\geq & |A(C^n)| \times 
(1+\zeta)^s. 
\end{eqnarray} 
Since $s$ can be made arbitrarily large, this means the $|A|(C^n)$ is
not bounded. \hfil $\qed$ \vskip0.3cm

\section{Discussion} \label{disc}

We have seen that for a class of finite dimensional systems the quantum vector pre-measure on $\fA$
does not in fact admit extensions to the sigma algebra $\fS_\fA$ generated by $\fA$. We now discuss
the implications of these results.

To start with, the lack of an extension does not by itself imply that no physical observables can be
constructed. For the finite unitary systems {\it all} the events in $\fA$ are measurable and hence
are physical observables. The lack of an extension simply means that while finite time questions are
observables, not all infinite time questions are.

For causal sets on the other hand, one is interested in covariant or label invariant observables.
Since the growth process generates only labelled causal sets, the events in $\fA$ are not themselves
observables.  One method of constructing covariant events \cite{observables} is to first generate
the labelled sigma algebra $\fSA$ and then take its quotient $\fS'$ with respect to
relabellings. $\fS'$ is a sigma algebra over the space of unlabelled past-finite causal sets
$\Omega'$, and is collection of covariant events.  For the classical stochastic growth models of
\cite{csg} since the extension of the probability pre-measure to $\fSA$ is guaranteed starting from
a given probability pre-measure on $\fA$ this procedure gives rise to a unique covariant probability
measure space $(\Omega', \fS', \mu')$, where $\mu'$ is a covariant measure. If a quantum vector
pre-measure extends to a vector measure on $\fS_\fA$, a similar procedure will give rise to a
collection of covariant quantum observables.

When $p$ is real, i.e., for real quantum percolation, the pre-measure does extend to $\fS_\fA$, and
hence covariant quantum observables can be constructed along the lines of \cite{observables}. In
particular, from the product form of the decoherence functional (Eqn (\ref{product})) it is trivial
to see that {\it all} the covariant observables of classical transitive percolation are also
observables for real quantum percolation. In \cite{observables} it was shown that the covariantly
defined and physically accessible {\sl stem sets} generate a sub-sigma algebra $\fS'_S \subset
\fS'$. A {\sl stem} $c$ in a causal set $C$ is a sub-causal set of $C$ which contains its own
past\footnote{A stem set is a discrete version of a {\sl past set} for which $S=J^-(S)$, where
  $J^-(x)$ denotes the set of events in the causal past of $x$ \cite{penhawel}.}.  The stem set,
$stem(c) \subset \Omega'$ is then the set of (unlabelled) causal sets which contains a stem
isomorphic to $c$. Since the sets of measure zero in transitive percolation are also sets of measure
zero in real quantum percolation, the results of \cite{observables} imply that for $p>0$, $\fS'_S$
generates $\fS'$ upto sets of measure zero. Thus, at least in this simple example of quantum
dynamics, one recovers a complete set of covariant quantum observables.

As we have seen, for complex percolation with $p$ not real, $\muv$ does not extend to $\fS_\fA$, and
hence the construction of covariant observables, if at all possible, requires a different approach.
As discussed in Section \ref{qvm} countable additivity of the measure requires an unconditional
convergence of the right hand side of Eqn (\ref{ucc}). It is this that implies bounded variation for
finite dimensional vector measures.  One option then is to require that the measure on $\fSA$
satisfy only a {\it conditional} convergence, with the conditionality determined by a preferred
ordering of events in $\fA$. The ``nested'' structure of the cylinder sets (Eqns (\ref{cylsubset})),
(\ref{cylint}) suggests a natural ordering on $\fA$ in terms of cardinality. For example, for a
countable disjoint union of cylinder sets, a conditional convergence can be easily defined with
respect to such an ordering. However, in order for this prescription to work in general, it is
important to ensure that such an ordering procedure is unique.

A simpler alternative is to use transitive percolation as a template for the quantum vector measure,
by first calculating the probabilities for transitive percolation and then complexifying them. In
other words, if $P(\alpha)$ is the probability for the event $\alpha \in \fS_\fA$ for transitive
percolation, then one defines $A(\alpha)$ to be an appropriately complexified version. As a concrete
example, consider the event $\alpha_{o} \in \fS_\fA$ corresponding to the existence of an element
which is to the past of all other elements\footnote{A causal set with an element to the past of all
  other elements is referred to as ``originary''.}. The probability of such an event for transitive
percolation is given by the Euler function
\begin{equation} 
\phi(q) =\prod_{i=1}^\infty (1-q^i).  
\end{equation} 
Since $\phi(q)$ is finite for all $q \in \bC, \,\,|q| < 1$, we can
define $D(\alpha_0,\alpha_0) \equiv |\phi(q)|^2$. Whether this gives
rise to a genuine finitely bi-additive quantum measure on $\fSA$ would
depend on the detailed nature of the complexification procedure
adopted. 

If successful, such prescriptions, though seemingly ad-hoc, would be in keeping with the attitude one
adopts in physics.  Namely, the failure of a quantum measure to have a ``mathematical extension''
does not mean that it cannot have a ``physical extension''.  This happens in physics all the time:
even though one often comes across non-convergent expressions, we can make sense of them by applying
a physically meaningful cutoff and using the limit as the cutoff is taken away to define the
quantity. We can do the same thing here: define limits only in a physically meaningful way. Even if
quantities are only conditionally convergent these ``conditions'' are physically determined.

On the other hand, the failure of bounded variation could also have implications for physical
predictions. In classical probability theory, we expect that a set of zero quantum measure is one
which almost surely does not happen or is {\sl precluded}. However, sets of zero quantum measure can
contain sets of non-zero quantum measure \cite{qmeasure, qcovers}. Without going into the details of
an interpretational framework (see \cite{alogic,ks}) it is reasonable to assume that any set
contained in a set of zero quantum measure is also precluded and hence almost surely will not
happen.

We now show that bounded variation is sufficient to ensure that not all elements of $\fA$ are
contained in a set of measure zero for a large class of systems, and hence are not precluded.

\begin{claim} 
  For an event algebra $\fA$ generated by cylinder sets constructed from finite length strings as in
  (\ref{defcyl}) and which satisfy (\ref{cylsubset}) and (\ref{cylint})\footnote{The term ``cylinder
    sets'' is also used to describe the sets that generate the event algebra for continuous times,
    for example for Brownian motion. These sets do not satisfy (\ref{cylsubset}) and
    (\ref{cylint}).}, if $\muv$ is of bounded variation, then not every $\alpha \in \fA$ can be
  contained in a set of measure zero.
\end{claim} 

\noindent {\bf Proof:} Assume the contrary, i.e., that every $\alpha \in \fA$, $\alpha \subset
\Omega$ is either of measure zero or contained in a set of measure zero. Let $\Gamma(n)$ denote the
set of length $n$ strings. For any $\tg \in \Gamma(n)$ define $\tG \equiv \Gamma(n) \backslash \tg$
and a disjoint union of cylinder sets $\alpha_{\tg} \equiv \bigcup_{\gamma \in \tG} \cyl(\gamma) \in
\fA$. Then there exists a $\beta_{\tg} \supseteq \alpha_{\tg}$ which is of quantum measure zero,
i.e., $\parallel\muv(\beta_\tg)\parallel = D(\beta_\tg, \beta_\tg)=0$. This means that
$\muv(\beta_\tg)$ is a zero vector in $\cH$, so that $\muv(\Omega)=\muv(\beta_\tg^c)$ where
$\beta_\tg^c \subseteq \cyl(\tg)$, by the additivity of $\mu_v$.  Thus, $\parallel
\muv(\beta_\tg^c)\parallel =1$. Now consider the following partition of $\Omega$. For each $\tg \in
\Gamma(n)$, express $\cyl(\tg)$ as the disjoint union $ \beta_\tg^c \bigcup (\cyl(\tg) \backslash
\beta_\tg^c)$, so that $\Omega$ can be expressed as the disjoint union
\begin{equation} 
\Omega = \bigcup_{\tg \in \Gamma(n)} \beta_\tg^c
\bigcup (\cyl(\tg) \backslash \beta_\tg^c). 
\end{equation} 
Then,  
\begin{equation}
|\muv|(\Omega) \geq   \sum_{\tg \in \Gamma(n)} \parallel \muv(\beta_\tg^c)\parallel = 2^n, 
\end{equation} 
for each $n$ and is hence unbounded. \hfill $\qed$ \vskip0.3cm

While this does not prove that bounded variation is a necessary condition, the following is an
example which is not of bounded variation, and for which {\it every} set in $\fA$ is contained in a
set of measure zero and hence is precluded. In other words, the only event in $\fA$ that does occur
is $\Omega$ itself!  This system belongs to the class of generic unitary systems studied in Section
\ref{finite} and hence is not of bounded variation. 

\begin{claim} For a two state system whose dynamics is determined by the unitary operator
\begin{equation} \label{nir} 
U=\frac{1}{\sqrt 2} 
\begin{pmatrix} 
 1 & i   \cr 
i & 1  
\end{pmatrix} \end{equation} every element in $\fA$ is contained
in a set of measure zero, except for $\Omega$.
\end{claim}

\noindent {\bf Proof:} 

For a two dimensional system, $s_i \in \{1,2\}$.  If $e_1,e_2$ are the orthonormal basis vectors of
$\cH_1$, the trivial evolution of this basis  (i.e., via the identity map) to time $t=m$ gives the
orthonormal basis vectors $e_1(m), e_2(m)$ of $\cH_m$.  Wlog, let the initial state $\psi=e_1 \in
\cH_1$.  For $U$ given by Eqn (\ref{nir}) the restricted evolution $\tpsi_\gamma$ for any $m$-length
truncated history $\gamma=(s_1, s_2, \ldots, s_m)$ is
\begin{equation}
\tpsi_\gamma=\hC_{\gamma} \psi = \biggl( \frac{1}{\sqrt 2}\biggr)^{m-1} (i)^f e_{s_m} \delta_{s_1,1}
\end{equation} 
where $f \in \{0,1, \ldots m-1 \}$ denotes the number of ``flips'', i.e., transitions from $0$ to
$1$ or $1$ to $0$ in $\gamma$. For example, if $\gamma=(0,1,0)$, then $f=2$.
For this choice of initial state, $\tpsi_\gamma$ is identically zero on all histories with
$s_1=2$, and moreover for $f$ even, $e_{s_m}=e_1(m)$ while for $f$ odd, $e_{s_m}=e_2(m)$.   
Thus, truncated histories with the same $m,f$ values have the same $\tpsi_\gamma$.

If $\Gamma(m)$ denotes the set of all $m$-length strings, then the  {\it unrestricted} evolution
$\tpsi_\Omega$ from $\psi$ is 
\begin{eqnarray}   \label{omegastate}
\tpsi_\Omega & = &  \sum_{\gamma \in \Gamma(m)} \tpsi_\gamma=\sum_{\gamma \in \Gamma(m)}
\hC_\gamma\psi \nonumber \\ 
&=& a_1(m)\,  e_1(m) + a_2(m)\, e_2(m).
\end{eqnarray}  
Clearly $\tpsi_\Omega$ is independent of $m$ and satisfies the normalisation $\langle \tpsi_\Omega,
\tpsi_\Omega\rangle =1$. Contributions to $a_1(m)$ come only from truncated histories with even $f$ and
those to $a_2(m)$ only from truncated histories with odd $f$. The number of strings with precisely
$f$ flips is ${m-1\choose f}$, and hence summing over even and odd $f$ respectively we find   
\begin{eqnarray} 
a_1(m) &=&  
\left(\frac{1}{\sqrt{2}}\right)^{m-1}\, \sum_{j=0}^{j_{max}}\, (-)^j \, {m-1 \choose 2j}
 \nonumber \\ 
a_2(m) & = &  i \,  
 \left(\frac{1}{\sqrt{2}}\right)^{m-1}\,  \sum_{k=0}^{k_{max}}\, (-)^k \,{m -1 
 \choose 2k+1},  
\end{eqnarray} 
where, for $m$ even $j_{max}=k_{max}=\frac{1}{2}(m-2)$ and for $m$ odd $j_{max}=\frac{1}{2}(m-1)$
and $k_{max}=\frac{1}{2}(m-3)$.

For $m=4q+1$ the binomial expansion for $(1+i)^{4q}$ gives  
\begin{eqnarray} 
a_1(4q+1)&= & \biggl(\frac{1}{\sqrt{2}}\biggr)^{4q} (-)^q
\,\,2^{2q}\\
a_2(4q+1)&=& 0.  
\end{eqnarray} 
Now, for every $\gamma \in \Gamma(4q+1)$ whose first entry is $s_1=1$ and which has precisely $2q$
flips, the last entry $s_{4q+1}=1$. Hence the strings with $2q$ flips contribute only to
$a_1(4q+1)$, each with an amplitude $(\frac{1}{\sqrt{2}})^{4q}\,(-)^q$. Moreover, the
number of truncated histories with precisely $f=2q$  $\Gamma_1 \subset \Gamma(4q+1)$ is  
${4q \choose 2q}$. Since ${4q \choose 2q} > 2^{2q}$, it is therefore possible to pick a subset
$\Gamma_2$ of $\Gamma_1$ so that 
\begin{equation}\label{stuffed} 
\tpsi_\Omega = \sum_{\gamma \in \GF} \hC_{\gamma} \psi 
\end{equation} 
The complement of the set  $\bigcup_{\gamma \in \GF} \cyl(\gamma)$ is therefore of measure zero. 

We now show that every cylinder set $\cyl(\tg)$ with $\tg=(s_1, s_2, \ldots,
s_{\tilde m})$ contains an event $\alpha$ such that its complement is of zero measure. 
Since every event except $\Omega$ is in the complement of some cylinder set, it also belongs to the
complement of $\alpha$ which suffices to prove our result.

Let $\tf$ be the number of flips in $\tg$. Any event which is a subset of $\cyl(\tg)$ shares the
first $\tilde m$ entries with $\tg$ and hence has at least $\tf$ flips.  Choose a $q$ such that
$2q>\tf$ and let $\Gamma_{\tg} \subset \cyl(\tg)$ be the set of truncated histories of length $4q+1$
with precisely $2q$ flips. The cardinality of this set is ${4q-\tilde m \choose 2q -\tf}$ and each
truncated history in it contributes a factor of $(\frac{1}{\sqrt 2})^{4q} \,(-)^q$ to $a_1(4q+1)$.
Since $\tf \in \{0, 1, \ldots, \tilde m\}$, if $q \geq 3$
\begin{equation} 
{4q-\tilde m \choose 2q -\tf} \geq  {4q-\tilde m \choose 2q -\tilde m} > {3q \choose
  q} \geq 2^{2q}.  
\end{equation} 
Thus, there exists a $\Gamma_2 \subset \Gamma_{\tg} \subset \Gamma_1$ for which Eqn (\ref{stuffed})
is satisfied so that the complement of  the event  $\bigcup_{\gamma \in \Gamma_2}$ is of measure
zero. Thus, every event in the complement of $\cyl(\tg)$ is contained in a set of measure zero.  

Since this is true for all cylinder sets, it means that every event in $\fA$ is contained in a set
of measure zero.  \hfill $\qed$ \vskip0.3cm

It is the purpose of this work to lay some of the groundwork for future investigations into
physically realistic examples for which the histories Hilbert space is infinite dimensional. These
include the Schrodinger particle and the quantum random walk \cite{qrw,hilbert}. The existence of an
extension of the quantum vector pre-measure in this infinite dimensional case requires a weaker
condition than bounded variation, but there are indications that even this is not satisfied for the
Schrodinger particle \cite{geroch}. This may suggest that time of passage questions, notoriously
difficult to pose in the standard approach to quantum theory, are not observables even within the
quantum measure approach.

\vskip0.3cm

\noindent{{\bf Acknowledgements:}} We thank Rafael Sorkin and Bob
Geroch for useful discussions. This work was supported in part by the
Royal Society International Joint Project 2006-R2.

\end{document}